\documentclass[conference]{IEEEtran}
\IEEEoverridecommandlockouts

\usepackage{cite}
\usepackage{amsmath,amssymb,amsfonts}
\usepackage{algorithmic}
\usepackage{graphicx}
\usepackage{textcomp}
\usepackage{xcolor}
\usepackage{array,multirow}
\usepackage{tablefootnote}
\usepackage{threeparttable}
\usepackage{relsize}
\usepackage{enumitem}
\usepackage{amsmath}
\usepackage{upgreek}
\usepackage[caption=false, font=footnotesize]{subfig}
\usepackage{url}

\def\BibTeX{{\rm B\kern-.05em{\sc i\kern-.025em b}\kern-.08em
    T\kern-.1667em\lower.7ex\hbox{E}\kern-.125emX}}

\makeatletter
\IEEEtriggercmd{\reset@font\normalfont\fontsize{7.9pt}{8.60pt}\selectfont}
\newcommand*\titleheader[1]{\gdef\@titleheader{#1}}
\AtBeginDocument{%
  \let\st@red@title\@title
  \def\@title{%
    \bgroup\normalfont\normalsize\centering\@titleheader\par\egroup
    \vskip1ex\st@red@title}
}
\makeatother
\IEEEtriggeratref{1}

\titleheader{\color{gray}\vspace{-30pt}\hspace{-3pt}Presented at the 30th IFIP/IEEE Int'l. Conference on Very Large Scale Integration (VLSI-SoC), 03/10--05/10 2022, Patras, GR.\\This is the authors’ version of the paper. The final published version is posted on IEEE Xplore.\\ DOI: \url{https://doi.org/10.1109/VLSI-SoC54400.2022.9939566}}
    
    \title{Towards Employing FPGA and ASIP Acceleration to Enable Onboard AI/ML in Space Applications\thanks{Work partially supported by the CAIRS21 research activity of the European Space Agency (ESTEC num. 4000135491/21/NL/GLC/ov).\\[-20pt]}}

\makeatletter
\def\ps@IEEEtitlepagestyle{
  \def\@oddfoot{\mycopyrightnotice}
  \def\@evenfoot{}
}
\def\mycopyrightnotice{
  {\footnotesize
  \begin{minipage}{\textwidth}
  \centering\color{gray}%
  ~\copyright~2022 IEEE.  Personal use of this material is permitted.  Permission from IEEE must be obtained for all other uses, in any current or future media, including reprinting/republishing this material for advertising or promotional purposes, creating new collective works, for resale or redistribution\\to servers or lists, or reuse of any copyrighted component of this work in other works.
  \end{minipage}
  }
}
    
\begin{document}

\author{\IEEEauthorblockN{Vasileios Leon\IEEEauthorrefmark{1}, 
    George Lentaris\IEEEauthorrefmark{1},
    Dimitrios Soudris\IEEEauthorrefmark{1},
    Simon Vellas\IEEEauthorrefmark{2},
    Mathieu Bernou\IEEEauthorrefmark{2}}\\[-11pt]
    \IEEEauthorblockA{\IEEEauthorrefmark{1}\emph{National Technical University of Athens, 15780 Athens, Greece}}
    \IEEEauthorblockA{\IEEEauthorrefmark{2}\emph{OHB-Hellas, 15124 Marousi, Greece}}
    }
    
\maketitle

\IEEEpubidadjcol

\begin{abstract}
The success of AI/ML in terrestrial applications and the commercialization of space are now paving the way for the advent of AI/ML in satellites. However, the limited processing power of classical onboard processors drives the community towards extending the use of FPGAs in space with both rad-hard and Commercial-Off-The-Shelf devices. The increased performance of FPGAs can be complemented with VPU or TPU ASIP co-processors to further facilitate high-level AI development and in-flight reconfiguration. Thus, selecting the most suitable devices and designing the most efficient avionics architecture becomes crucial for the success of novel space missions. The current work presents industrial trends, comparative studies with in-house benchmarking, as well as architectural designs utilizing FPGAs and AI accelerators towards enabling AI/ML in future space missions.
\end{abstract}


\bstctlcite{IEEEexample:BSTcontrol}

\section{Introduction}

The emergence of powerful embedded accelerators
and novel
Artificial Intelligence (AI) / Machine Learning (ML)
functionalities
is revolutionizing the terrestrial applications. 
However, 
the space industry 
is still striving to adopt these new technologies for onboard computing
due to historical and technical reasons \cite{ai_space}.
Classic onboard processors
cannot cope with
the increased demands of space applications
for high I/O bandwidth 
and processing throughput.
Hence, 
the space industry is forced
to examine 
Field-Programmable Gate Arrays (FPGAs)
and
Application-Specific Instruction-set Processors (ASIPs),
such as 
the Vision Processing Units (VPUs)  
and Tensor Processing Units (TPUs).
The onboard use of these accelerators
would offer high performance,
while bringing
``AI/ML one step closer to space''.
Towards improving the
Size, Weight, Power \& Cost (SWaP-C)
and providing heterogeneous computing,
the community 
also studies mixed-criticality architectures for
space avionics \cite{mixed_criticality}.
In this context,
depending on mission specifics,
the architecture
consists of more elaborate combinations 
of radiation-hardened and Commercial-Off-The-Shelf (COTS)
components.

Among others, the FPGAs
are being evaluated as 
accelerators of Vision-Based Navigation (VBN)
\& Earth Observation (EO) functions
or framing processors (handling of the instruments/sensors).
Research has been conducted
on both
radiation-hardened 
(e.g., NanoXplore's BRAVEs)
\cite{ahs_brave, access_brave, shyloc_fpga}
and
COTS
(e.g., Xilinx's Zynq-7000s and Zynq UltraScale MPSoCs)
\cite{mpsoc_vbn, mpsoc_cnn}
devices. 
Regarding ASIPs,
Intel's Myriad VPUs
have attracted great interest,
as they excel in low-power embedded computer vision and AI. 
These COTS System-on-Chips (SoCs)
are mainly used as onboard accelerators \cite{leotome, icecs_hpcb, fsat_myriad, myriadx_space} in mixed-criticality co-processing architectures. 
More recently,
Google's Edge TPU has gained momentum 
for accelerating Convolutional Neural Networks (CNNs),
and it is already considered for space avionics
\cite{tpu_spacecube}. 

In this work,
we focus on 
mixed-criticality architectures 
utilizing FPGA and AI ASIP devices.
We aim to facilitate the onboard execution of demanding AI/ML algorithms, 
provide HW virtualization, 
and
meet the dependability requirements of space missions.
The goal of the current paper is versatile:
\begin{itemize}
    \item To present today's landscape of the space industry's market and highlight the requirements of novel space applications and future missions. 
    \item To consider different variants of co-processing architectures integrating
    both radiation-hardened and COTS devices:
    microcontrollers, FPGAs, VPUs and/or TPUs.   
    \item To consider the programming model and implementation techniques on such FPGA- and ASIP-based architectures.
    \item To report preliminary experimental results from the implementation of CNNs and computer vision kernels on the FPGA and ASIP accelerators. 
\end{itemize}

\section{Background}

\subsection{The Market of Space Industry}
Following the space developments during the past decade, a clear change 
can be observed. 
Reduced launch costs have made space more accessible,
and low-cost, modular CubeSats can be built and financed even by universities. 
The use of lower-cost commercial HW is reducing the risk of failure, inviting many new actors to join the sector, driving innovation and demonstrating novel applications and mission scenarios.

At the same time, the market itself is calling for innovation in traditional satellite design approaches in order to cover modern application demands and address pressing societal issues (e.g., security, crisis management).
State-of-the-art large institutional EO missions are producing massive amounts of data, often up to 20 Gbit/s, 
introducing major challenges in the data-handling systems. 
On the other hand, there is an increasing need for near real-time, low-latency data availability in several time-critical applications, 
e.g., 
in the domains of emergency and natural disaster response, border monitoring and security.

The very high demands for fast processing and data availability
are the driving forces behind the search for new approaches in satellite data processing architectures. 
Conservative, ultra-high reliability approaches of the past are being replaced by efficient, modular and flexible 
(in HW and SW) high-performance HW implementations.
Concurrently, successful data-processing approaches used in terrestrial applications are being examined for space. 
Most notably, AI is proven to be an invaluable solution to the needs of modern satellites,
allowing for increased data value, reduced latency, increased system autonomy, reduced cost and new mission scenarios.

Based on the analysis of the EO market needs in the 2022 EUSPA EO Market Report \cite{euspa_eo}, we analyzed the markets 
that are expected to grow the most in the next 10 years in EU (Agriculture, Emergency management, Maritime, Environmental),
and we selected the most relevant to benefit from AI-based onboard processing. From all markets, 27 applications were identified, 60\% of which are expected to require or exploit the onboard AI functionalities. 

Finally, we have gained valuable insight 
by evaluating 
more than 30 providers of high performance onboard processing solutions including SoCs, CPUs, FPGAs, AI accelerators as well as complete computer solutions. A clear trend is evident in the rise of AI-capable HW in the form of FPGAs and specialized AI computational units, 
whereas modular, mixed-criticality implementations are necessary to address the extreme data processing needs. In any case, alongside novel HW architectures, reliable and flexible HW is needed to support these platforms. Currently, around 25\% of the identified onboard solution providers focus on SW, with $\sim$60\% of these offering complete and mature HW and SW solutions.

\subsection{FPGAs in Space}
FPGAs are integrated in space avionics architectures
mainly for 
accelerating high-performance DSP/AI algorithms
\cite{mpsoc_vbn, mpsoc_cnn, ahs_brave, access_brave},
performing fast data transformations 
(e.g., compression) \cite{shyloc_fpga},
and
handling the space-qualified sensors/instruments 
(e.g., SpaceWire and SpaceFibre) \cite{leotome}. 
There are several industrial FPGA-based boards, 
such as 
Xiphos' Q8S, 
Innoflight's CFC-400 
and Trenz Electronic's TE08XX
(all integrating Xilinx's Zynq UltraScale+ MPSoC),  
and
Cobham Gaisler's GR-VPX-XCKU060
(featuring the Kintex UltraScale XCKU060). 

In more detail, 
the work in \cite{mpsoc_vbn}
proposes
a re-configurable SW/HW onboard processor 
for VBN on MPSoC.
The authors of \cite{mpsoc_cnn} 
deploy Xilinx's DPU on MPSoC
to provide onboard CNN inferencing 
for spacecraft pose estimation.
Regarding the European space-grade FPGAs,
NG-Medium and NG-Large are tested
as onboard high-performance DSP accelerators
in \cite{ahs_brave} and \cite{access_brave},
respectively. 
The same devices
are evaluated among other
radiation-hardened/tolerant FPGAs
for implementing 
the 
SHyLoC 2.0 CCSDS 
121 \& 123 
lossless 
compression standards \cite{shyloc_fpga}. 
Finally, 
in the co-processing architecture of \cite{leotome},
the FPGA
is responsible for receiving the data 
via SpaceWire and performing the required transcoding.

\subsection{ASIPs in Space}
COTS ASIP accelerators,
such as GPUs, VPUs, and TPUs, 
are constantly being evaluated
for accelerating DSP \cite{gpu, leotome, icecs_hpcb} 
and AI \cite{icecs_hpcb, fsat_myriad, myriadx_space, tpu_spacecube} kernels.
As a result,
various relevant space products have started to appear 
in the market. 
Aitech's S-A1760 Venus,
which integrates the Jetson TX2i GPU,
is a space-rated rugged
AI processor for short-duration spaceflights. 
Ubotica's CogniSat is a CubeSat board
that exposes the Myriad2 VPU to the payload developer
for inferencing AI and computer vision algorithms.
In the same context,
NASA's SC-LEARN is 
a SpaceCube SmallSat Card
with three TPUs, 
targeting autonomous operations and onboard AI acceleration.

In terms of related works,
the authors of \cite{gpu}
examine the applicability of GPUs in the space domain
by analyzing algorithms and workloads of space applications
and benchmarking various devices.
The Myriad2 VPU is used to accelerate a sophisticated 
computer vision pipeline for estimating Envisat's pose \cite{leotome}. 
The same VPU 
is the main DSP/AI accelerator in the co-processing architecture of \cite{icecs_hpcb}.
Myriad2 is also equipped onboard 
as CNN demonstrator for EO
in the
$\mathrm{\Phi}$-{S}at-1 CubeSat mission of ESA \cite{fsat_myriad}.
Furthermore, 
in \cite{myriadx_space},
the MyriadX VPU is used for implementing neural networks classifiers,
which are trained on Mars imagery from the Reconnaissance Orbiter and Curiosity rover.
Regarding TPUs,
the work of \cite{tpu_spacecube}
reports preliminary results 
for hyperspectral image classification using 
an 1D multi-layer perceptron 
and a spectral-spatial CNN.

\section{Architecture Overview and Technical Details}

\subsection{FPGA \& ASIP Co-Processing Architecture}

The main purpose of our payload computer is 
to accelerate demanding AI/ML together with classical DSP algorithms,  
to provide some level of HW virtualization 
in order to facilitate SW upload and multi-user service,
as well as to meet the increased dependability requirements of space missions,
all with a limited budget of power \& size.
To this end, we consider a mixed-criticality architecture
consisting of 3 key HW components (regardless of component-level redundancy):
\begin{itemize}
\item a space-grade CPU (e.g., GR712) 
\item a COTS SoC FPGA (e.g., Zynq UltraScale+ MPSoC)
\item a COTS ASIP AI accelerator (e.g., TPU/VPU)
\end{itemize}

The space-grade CPU serves as the supervisor of the entire system 
and guarantees that the satellite's data
will not be lost in case of COTS failures during flight. 
That is to say,
on one hand,
the supervisor should promptly correct/restart the failed blocks  
and gradually restore the normal operational mode of the system.
On the other hand,
in case of certain permanent failures, 
the space-grade CPU should begin handling alternative route/store/preprocessing operations
of the incoming data,
and hence,
allow the system to revert from a high- to low-performance mode
(to offer some minimum functionality instead of zero response, i.e., complete data loss).

The role of the COTS SoC FPGA is threefold.
First, 
the FPGA will host heritage space IPs/accelerators, e.g., SHyLoC compression,
which can considerably improve the development time of new space avionics products.
Second,
it will facilitate tightly-coupled HW/SW co-processing for DSP and/or AI functions; 
the embedded processor increases greatly the flexibility of the FPGA
by undertaking inefficiently-accelerated SW parts 
(executing them closely next to the HDL parts)
or even enabling virtualization techniques at OS level.
Third, the FPGA can implement a variety of high-rate protocols,
e.g., SpaceWire, and interface 
with multiple instruments in a concurrent and deterministic fashion.
Moreover, these protocols can be modified during the development phase and
allow the engineers to adapt/reuse the proposed system 
for distinct satellites with minimum time-to-market. 
We note that the dependability of the COTS FPGA can increase 
to a certain level via soft error mitigation techniques 
(e.g., triple modular redundancy, memory scrubbing, watchdogs)
depending on mission requirements and design trade-offs.

The role of the AI/ML ASIP will be to execute 
the most computationally demanding parts of the application
in the most efficient fashion (given its specialized HW). 
In AI-dominated applications,
employing the TPU/VPU chips could significantly benefit 
the overall performance and performance-per-Watt of the system.
Secondarily, utilizing such accelerators enables
3rd party users to develop the SW for the proposed system with less programming effort/time
by leveraging popular commercial development frameworks.
We note that the dependability of the ASIP component
is not necessarily as crucial as that of the FPGA.

The aforementioned architecture is a relatively generic design
relying on few key components/ideas,
such as SoC and ASIP customization.
The efficiency of its implementation also relies 
on board/chip specifics, interfaces, and targeted mitigation techniques.
Our work is to explore different possibilities/combinations
and assess the best configuration per mission.
In particular,
we need to assess the trade-offs 
between Zynq UltraScale+ MPSoC \cite{trenz}
and XKCU060 \cite{vpx}
for the given FPGA role
or even the trade-off in using rad-hard BRAVE technology \cite{nx}.
Furthermore,
the aforementioned FPGA can connect with USB ASIP accelerators
(USB TPU \cite{tpuusb} and NCS2 VPU \cite{vpuusb}) or with CogniSat over Ethernet \cite{cognisat}.
Finally,
combining fewer components is still an option for lower-cost missions,
e.g., omitting the space-grade CPU.

\subsection{Development on FPGA and ASIP}

\subsubsection{TPU}
The Edge TPUs
rely on a systolic array of multipliers and accumulators
and an on-chip SRAM for storing
the model's executable and parameters.
There are both TPU development boards and USB accelerators. 
The boards feature 
ARM processors, 
I/O interfaces, 
and various memories 
(flash and DDR).
In terms of development, 
the TPUs support 
TensorFlow models,
which need to be quantized and converted to TensorFlow Lite, 
and then compiled via the Edge TPU compiler.
The 8-bit integer quantization
can be performed either using the TensorFlow quantization-aware training or post-training quantization. 
The inference requires the Edge TPU runtime, and can be performed via the Python and C++ APIs of TensorFlow Lite.

To build a network for the TPU,
the developers follow the typical TensorFlow design flow,
however, 
they need to be aware of the operations supported in TPU
(e.g., 3D convolutions are not supported). 
The Edge TPU compiler offers various features than can be exploited towards increased performance and/or reliability.
The co-compilation of models enables
to cache them inside TPU 
avoiding the timing penalties to run a new network.
Moreover, 
the compiler can divide the model into
segments and
parallelize its execution to multiple TPUs.

\subsubsection{VPU}
The Myriad VPUs offer
two general-purpose LEON4 CPUs,
multiple SIMD \& VLIW programmable cores,
and HW imaging filters.
MyriadX also features 
a dedicated AI accelerator engine. 
Their heterogeinity extends to the memory hierarchy,
which includes 
on-chip DDR DRAM,
scratchpad SRAM,
and caches. 
The developers use the MDK design suite for 
implementing custom DSP/AI kernels on the programmable vector cores and filters.
Moreover,
the OpenVINO toolkit is used
for fast CNN deployment on the AI engine of MyriadX,
either on the SoC or the NCS2 USB accelerator.

For custom development,
we share the workload among
the vector cores
using either static
or dynamic task scheduling
(task = processing of an image stripe). 
The dynamic scheduling is favored 
in content-dependent workloads. 
We also apply workload-specific
memory configuration
with respect to the memory accesses and data sizes. 
At lower level,
we apply wordlength tuning
(VPUs support various arithmetic formats),
and opt to 
reduce the number of buffers
due to the limited size of the scratchpad memory. 
To enable SIMD functionalities,
we arrange accordingly the data 
and call the respective vendor's routines.
Specifically for CNNs,
OpenVINO
supports models from various frameworks (e.g., TensorFlow),
and allows to tune the network's frozen graph via its model optimizer
(e.g., apply 
linear operation fusing, grouped convolution fusing, network pruning). 

\subsubsection{BRAVE}
The space-grade BRAVE FPGAs 
offer all the traditional logic-programmable resources,
which are 
organized in rows dedicated to 
functional elements, RAMBs, or DSPs. 
Moreover, 
NG-Large and NG-Ultra include ARM processors,
i.e., 
the single-core Cortex-R5 
and the quad-core Cortex-R52,
respectively.
The BRAVE FPGAs also provide features that are essential in onboard computing, 
such as the high-performance SpaceWire interface
and memory scrubbing.
In terms of software tools,
the NXmap design suite supports all the typical FPGA design stages.
The developers build the projects with Python scripts,
where they can configure their designs using a wide range of functionalities and settings. 
For the efficient implementation of demanding
computer vision kernels,
which rely on convolutions and memory accesses similar to CNNs, 
we exploit every tool setting.
For balancing the resource utilization,
we use the routines for custom mapping
(e.g., {\fontsize{9.5}{10.5}\selectfont\texttt{addmappingDirective}})
to explicitly specify the FPGA block
for implementing our arithmetic and memory components. 
Similarly,
we experiment with the general tool settings
(e.g., 
{\fontsize{9.5}{10.5}\selectfont\texttt{DensityEffort}}, 
{\fontsize{9.5}{10.5}\selectfont\texttt{CongestionEffort}}, 
{\fontsize{9.5}{10.5}\selectfont\texttt{PolishingEffort}})
and specify custom regions on the floorplan 
(e.g., 
{\fontsize{9.5}{10.5}\selectfont\texttt{createRegion}}, 
{\fontsize{9.5}{10.5}\selectfont\texttt{addRAMLocation}}),
towards achieving a better clock frequency.

\section{Preliminary Evaluation} 

\begin{figure}[!b]
\includegraphics[width=\columnwidth]{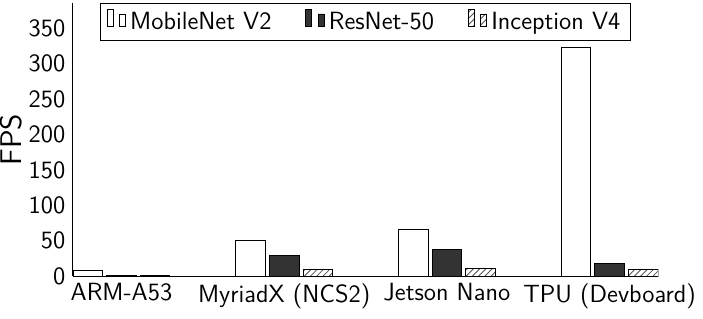}
    \vspace*{-23pt}
    \caption{Performance of Embedded Devices for CNN Inference}
    \label{comp}
\end{figure}

Fig. \ref{comp} reports the throughput
of the embedded accelerators for inferencing CNNs
of distinct complexity,
while
Table \ref{tb_ai} reports results from the implementation of a custom CNN kernel for ship detection.
Finally,
Table \ref{tb_cv} reports results
from the implementation of computer vision algorithms,
i.e., a sophisticated pipeline for satellite pose estimation
and hardware kernels for feature detection
(edges and corners)
and depth extraction
(stereo vision, 3D scene reconstruction).

Regarding AI acceleration,
TPU outperforms all the other accelerators
for small CNNs (MobileNet V2 and ShipDetect).
In particular,
it delivers
5--8$\times$ more FPS than
the Myriad VPUs and the Jetson Nano GPU,
while
compared to Zynq-7020,
it provides a speedup of 
1.4$\times$. 
For CNNs of higher complexity.
the results vary.
For ResNet-50, 
MyriadX and Jetson Nano exhibit $\sim$2$\times$
throughput versus TPU,
while for Inception V4,
all the accelerators
sustain $\sim$10 FPS.
Finally, ARM-A53 delivers $\sim$0.5 FPS for the demanding CNNs,
proving the inefficiency of CPUs for such compute-intensive tasks.

\begin{table}[!b]
\renewcommand{\arraystretch}{1.1}
\setlength{\tabcolsep}{2.1pt}
\caption{Results for ShipDetect CNN (128$\times$128$\times$3 Img., 132K Par.)}
\vspace{-5pt}
\label{tb_ai}
\centering
\begin{tabular}{c|ccccc}
\hline 
  & \textbf{ARM-A53}  &\textbf{Myriad2} & \textbf{Jetson Nano} & \textbf{Zynq-7020} & \textbf{TPU}\\[-1pt]
    & Devboard  & MVboard & Board & Zybo & DevBoard\\
 \hline \hline 
\textbf{FPS}              & 45 & 125 & 204 & 730 & 1000\\
\textbf{Watt}             & 4 & 1 & 10 & 5 & 5 \\
\textbf{FPS-per-Watt}    & 11 & 125 & 20 & 146 & 200\\
\hline
\end{tabular}
\end{table}

\begin{table}[!b]

\renewcommand{\arraystretch}{1.1}
\setlength{\tabcolsep}{6.5pt}
\caption{Results for Computer Vision Algorithms (1024$\times$1024 Img.)}
\vspace{-5pt}
\label{tb_cv}
\centering
\begin{tabular}{lc|ccc}
\hline 
\multicolumn{1}{c}{\textbf{Algorithm}}  & \textbf{Device}  &\textbf{FPS} & \textbf{Watt} & \textbf{FPS-per-Watt} \\
 \hline \hline 
Pose Estimation       & Zynq-7020  & 12  & 5 & 2.4 \\
Pose Estimation       & Myriad2   & 4   & 1 & 4 \\
Feature Detection     & NG-Large   & 10  & 4 & 2.5 \\
Depth Extraction      & NG-Large   & 0.2 & 4 & 0.05 \\
\hline
\end{tabular}
\end{table}

For traditional computer vision tasks,
Zynq-7020 is the best solution for high-performance 
($\sim$12 FPS on average),
while Myriad2 
provides the most power-efficient solution,
i.e., $\sim$4 FPS at 1W.
On the other hand,
the NG-Large BRAVE FPGA
provides sufficient throughput for
standalone computer vision kernels.
Given that most
VBN applications require 1--10 FPS,
the throughput for feature detection
is sufficient, 
allowing other functions of algorithmic pipelines 
in space applications
to finish on time. 
Depth extraction for 3D scene reconstruction
is performed every $\sim$7s with NG-Large
on image pairs of 1024$\times$1024 pixel-size. 

\section{Conclusion}
After a market analysis,
we considered a generic mixed-criticality architecture 
for payload processing in space,
which combines space-grade CPU
with COTS FPGA and ASIP AI acceleration.
We discussed variants of the architecture, 
and based on lab equipment for early prototyping purposes, 
we reported technical details including preliminary evaluation results 
for the key candidate components (FPGA, VPU, TPU).

\linespread{0.95}

\bibliographystyle{IEEEtran}
\bibliography{REFERENCES.bib}

\end{document}